\documentclass[usenatbib]{mn2e}
\usepackage{graphicx,tabularx,amsmath,amssymb,upgreek,wasysym,mathtools,multirow}
\numberwithin{equation}{section}
\title[Are the majority of Sun-like stars single?]{Are the majority of Sun-like stars single?}
\author[A. P. Whitworth and O. Lomax]{A. P. Whitworth\thanks{E-mail: ant@astro.cf.ac.uk} and O. Lomax\\
School of Physics and Astronomy, Cardiff University, Cardiff CF24 3AA, Wales, UK}

\begin{document}
\pagerange{\pageref{firstpage}--\pageref{lastpage}} \pubyear{2013}
\maketitle
\label{firstpage}

%%%%%%%%%%%%%%%%%%%%%%%%%%%%%%%%%%%%%%%%%%%%%%%%%%%%%%%%%%%%%%%%%%%%%%%%%%%%%%%%
\begin{abstract} 
It has recently been suggested that, in the field, $\sim\!\!56\%$ of Sun-like stars ($0.8\,{\rm M}_{_\odot}\la M_\star\la 1.2\,{\rm M}_{_\odot}$) are single. We argue here that this suggestion may be incorrect, since it appears to be based on the multiplicity frequency of systems with Sun-like primaries, and therefore takes no account of Sun-like stars that are secondary (or higher-order) components in multiple systems. When these components are included in the reckoning, it seems likely that only $\sim\!46\%$ of Sun-like stars are single. This estimate is based on a model in which the system mass function has the form proposed by Chabrier, with a power-law Salpeter extension to high masses; there is a flat distribution of mass ratios; and the probability that a system of mass $M$ is a binary is $\,0.50 + 0.46\log_{_{10}}\!\left(M/{\rm M}_{_\odot}\right)\,$ for $\,0.08\,{\rm M}_{_\odot}\leq M\leq 12.5\,{\rm M}_{_\odot}$, $\,0\,$ for $\,M<0.08\,{\rm M}_{_\odot}$, and $\,1\,$ for $\,M>12.5\,{\rm M}_{_\odot}$. The constants in this last relation are chosen so that the model also reproduces the observed variation of multiplicity frequency with primary mass. However, the more qualitative conclusion, that a minority of Sun-like stars are single, holds up for virtually all reasonable values of the model parameters. Parenthetically, it is still likely that the majority of {\it all} stars in the field are single, but that is because most M Dwarfs probably are single.
\end{abstract}
%%%%%%%%%%%%%%%%%%%%%%%%%%%%%%%%%%%%%%%%%%%%%%%%%%%%%%%%%%%%%%%%%%%%%%%%%%%%%%%%


%%%%%%%%%%%%%%%%%%%%%%%
\section{Introduction}%
%%%%%%%%%%%%%%%%%%%%%%%

The multiplicity statistics of Sun-like stars in the field have recently been re-evaluated by \citet{Raghetal2010}, using a volume-limited sample of 454 stars within $25\,{\rm pc}$. They estimate that, for systems having a Sun-like primary, the multiplicity frequency is $m_{_{\rm S}}(M_{_1}\!=\!{\rm M}_{_\odot})\!=\!0.44\pm0.02$.\footnote{Here the subscript {\sc s} records the fact that $m_{_{\rm S}}(M_{_1}\!=\!{\rm M}_{_\odot})$ is a system-property. $\,``(M_{_1}\!=\!{\rm M}_{_\odot})"\,$ is to emphasise that -- for practical observational reasons -- the convention is to estimate $m_{_{\rm S}}$ as a function of primary mass, $M_{_1}$, rather than system mass. We use $m_{_{\rm S}}$ for multiplicity frequency, and not $mf$ as is normal, because mathematical expressions involving a double symbol like $mf$ are inevitably confusing.} 

The multiplicity frequency for systems having a primary of mass $M_{_1}$ is defined as
\begin{eqnarray}
m_{_{\rm S}}(M_{_1})&=&\frac{B+T+Q+...}{S+B+T+Q+...}\,,
\end{eqnarray}
where $S$ is the number of single stars of mass $M_{_1}$, $B$ is the number of binary systems having a primary of mass $M_{_1}$, $T$ is the number of triple systems having a primary of mass $M_{_1}$, $Q$ is the number of quadruple systems having a primary of mass $M_{_1}$, and so on. Implicit in the above statement is the fact that in reality one must consider a finite interval of mass, in order to have meaningful statistics.

Thus, the \citet{Raghetal2010} estimate of $m_{_{\rm S}}({\rm M}_{_\odot})$ indicates that the number of single Sun-like stars exceeds, by a factor of $1.28\pm0.08$, the number of Sun-like stars that are primaries in multiple systems. This is not quite the same as the inference made by \citet{Raghetal2010}, {\it ``the majority ... of solar-type stars are single''}, and reiterated by \citet{DuchKrau2013} {\it ``a slight majority of all field solar-type stars are actually single''}, since it takes no account of Sun-like stars that are secondaries, tertiaries, etc. in multiple systems, only those that are primaries.

In the sequel we show that the majority of Sun-like stars may, in fact, be in multiples. To keep the analysis simple we consider only single and binary systems, so we are concerned with evaluating the number of Sun-like stars that are secondaries in binary systems. Consideration of higher-order multiple systems would only strengthen our conclusion by bringing into the reckoning additional components that might be Sun-like. Section \ref{SEC:MODEL} presents the model of binary statistics that we use, the ranges in which we allow the model parameters to vary, and the basic analysis. Section \ref{SEC:RES} presents the results, i.e. the fractions of Sun-like stars that are single, primary or secondary, and how these fractions depend on the parameter choices made. Section \ref{SEC:CONC} summarises our conclusions.

%%%%%%%%%%%%%%%%%%%%%%%%%%%%%%%%%%%%%%%%%%%%%%%%%%%%%%%%%
\section{Model parameters and analysis}\label{SEC:MODEL}%
%%%%%%%%%%%%%%%%%%%%%%%%%%%%%%%%%%%%%%%%%%%%%%%%%%%%%%%%%

In order to model the binary statistics of Sun-like stars and estimate what fraction are secondaries in binaries, we need to specify the distribution of system masses, the fraction of systems that are binaries (as a function of {\it system mass}), and the distribution of mass ratios in binary systems.

We define the system mass to be $M$, and the masses of stars to be $M_i$, where $i\!=\!0$ corresponds to a single star (hereafter a {\it single}), $i\!=\!1$ corresponds to the primary in a binary (hereafter a {\it primary}), and $i\!=\!2$ to the secondary in a binary (hereafter a {\it secondary}). In addition, we introduce the corresponding logarithmic variables
\begin{eqnarray}
\mu&=&\log_{_{10}}\!\left\{\!\frac{M}{{\rm M}_{_\odot}}\!\right\}\,,\hspace{1.0cm}{\rm (systems)}\,;\\
\mu_i&=&\log_{_{10}}\!\left\{\!\frac{M_i}{{\rm M}_{_\odot}}\!\right\}\,,\hspace{1.44cm} {\rm (stars)}\,.
\end{eqnarray}
Since
\begin{eqnarray}
M&=&{\rm M}_{_\odot}\,10^\mu\;\,=\;\,{\rm M}_{_\odot}\,{\rm e}^{\ell\mu}\,,
\end{eqnarray}
where $\ell\equiv\ln(10)$, we have
\begin{eqnarray}
\frac{dM}{d\mu}&=&\ell\,M\,.
\end{eqnarray}
The mass ratio of a binary system is $q\!=\!M_{_2}/M_{_1}$.

%%%%%%%%%%%%%%%%%%%%%%%%%%%%%%%%%%%%%%%%%%%%%%%
\subsection{The distribution of system masses}%
%%%%%%%%%%%%%%%%%%%%%%%%%%%%%%%%%%%%%%%%%%%%%%%

We assume that low- and intermediate-mass systems ($\mu\!<\!\mu_{_{\rm C}}$, see below) have a log-normal mass distribution, as proposed by \citet{Chabrier2005},
\begin{eqnarray}\label{EQN:SYSMASS_LN}
\frac{d{\cal N}}{d\mu}_{_{\rm LN}}&=&\frac{1}{(2\pi)^{1/2}\sigma_{_{\rm S}}}\,\exp\!\left\{-\,\frac{(\mu-\mu_{_{\rm S}})^2}{2\sigma_{_{\rm S}}^2}\right\}\,,
\end{eqnarray}
where the subscript {\sc ln} is for log-normal. \citet{Chabrier2005} estimates that $\mu_{_{\rm S}}\!=\!-0.60$ and $\sigma_{_{\rm S}}\!=\!0.55$; the normalisation in \citet{Chabrier2005} is different because he uses physical units. We assume that these values are accurate to $\pm 0.05$, i.e. $-0.65\leq\mu_{_{\rm S}}\leq -0.55$ and $0.50\leq\sigma_{_{\rm S}}\leq 0.60$.

For higher-mass systems, $M\!>\!M_{_{\rm C}}$, we adopt a power-law distribution of masses, with exponent $\alpha$, i.e.
\begin{eqnarray}
\left.\frac{d{\cal N}}{dM}\right._{_{\rm PL}}&=&\frac{K}{\ell\,{\rm M}_{_\odot}}\,\left(\!\frac{M}{{\rm M}_{_\odot}}\!\right)^{\!-\alpha}\,,
\end{eqnarray}
where the subscript {\sc pl} is for power-law. It follows that 
\begin{eqnarray}\label{EQN:SYSMASS_PL}
\left.\frac{d{\cal N}}{d\mu}\right._{_{\rm PL}}&=&K\,\exp\!\left\{-(\alpha-1)\ell\mu\right\}\,.
\end{eqnarray}
The default exponent is $\alpha\!=\!2.35$, as first estimated by \citet{Salpeter1955}, and we assume this is accurate to $\pm 0.35$, i.e. $2.00\!\leq\!\alpha\!\leq\!2.70$.

We require that the join between the two distributions be smooth, i.e. at $\mu_{_{\rm C}}$ the distributions (Eqns. \ref{EQN:SYSMASS_LN} \& \ref{EQN:SYSMASS_PL}) and their slopes should be equal, so
\begin{eqnarray}\label{EQN:muC}
\mu_{_{\rm C}}\!&\!=\!&\!\mu_{_{\rm S}}\,+\,(\alpha-1)\,\ell\,\sigma_{_{\rm S}}^2\,, \\
K\!&\!=\!&\!\frac{1}{(2\pi)^{1/2}\sigma_{_{\rm S}}}\,\exp\!\left\{(\alpha-1)\ell\mu_{_{\rm C}}\,-\,\frac{(\mu_{_{\rm C}}-\mu_{_{\rm S}})^2}{2\sigma_{_{\rm S}}^2}\right\}.\hspace{0.5cm}
\end{eqnarray}
With the default values of the model parameters, the join occurs at $\mu_{_{\rm C}}\!\simeq\! 0.34$, corresponding to a system mass of $M_{_{\rm C}}\!\simeq\!2.2\,{\rm M}_{_\odot}$.

%%%%%%%%%%%%%%%%%%%%%%%%%%%%%%%%%%%%%%%%%%%%%%%%%%%%%%%
\subsection{The fraction of systems that are binaries}%
%%%%%%%%%%%%%%%%%%%%%%%%%%%%%%%%%%%%%%%%%%%%%%%%%%%%%%%

Since the distribution of mass ratios appears to vary slowly -- if at all -- with primary mass in the range $0.2\,{\rm M}_{_\odot}\!\la\!M_{_1}\!\la\!2\,{\rm M}_{_\odot}$ \citep{Jansetal2012,Raghetal2010,ReggMeye2013}, the fraction of systems that are binaries must depend on system mass in a similar way to the dependence of multiplicity frequency on primary mass, but displaced to slightly higher masses. We therefore assume that the fraction of systems that are binaries is given by
\begin{eqnarray}\label{EQN:BINFRAC}
\beta(\mu)\!\!&\!\!=\!\!&\!\!\left\{\begin{array}{lr}
0\,,&\mu\leq-\,\beta_{_0}/\beta_{_1};\\
\beta_{_0}+\beta_{_1}\mu\,,&-\,\beta_{_0}/\beta_{_1}\!<\!\mu\!<\!(1-\beta_{_0})/\beta_{_1};\\
1\,,&\mu\geq(1-\beta_{_0})/\beta_{_1}.
\end{array}\right.
\end{eqnarray}
The justification for adopting this functional form is given in \S \ref{SEC:FITb}, where we also derive the default values of $\beta_{_0}$ and $\beta_{_1}$, and their ranges, by fitting the observed run of multiplicity frequency, $m_{_{\rm S}}$ against primary log-mass, $\mu_{_1}$ (see Fig. \ref{FIG:mfvsM1}).

%%%%%%%%%%%%%%%%%%%%%%%%%%%%%%%%%%%%%%%%%%%%%
\subsection{The distribution of mass ratios}%
%%%%%%%%%%%%%%%%%%%%%%%%%%%%%%%%%%%%%%%%%%%%%

Estimates of mass ratios are sufficiently uncertain, and the samples subject to such severe selection effects, that the appropriate distribution is poorly constrained. A convention has emerged \citep[e.g.][]{ReggMeye2013,DuchKrau2013} that -- at least in the first instance -- the distribution should be fitted with a power-law (exponent $\gamma$), possibly with a minimum mass ratio ($q_{_{\rm MIN}}$). We therefore adopt the distribution
\begin{eqnarray}\label{EQN:MASSRATIO}
\frac{dP}{dq}&=&\left\{\begin{array}{ll}
0\,,&q<q_{_{\rm MIN}}\,;\\
(\gamma+1)\left(1-q_{_{\rm MIN}}^{(\gamma+1)}\right)^{-1}\,q^\gamma\,,\hspace{0.3cm}&q>q_{_{\rm MIN}}\,.\\
\end{array}\right.
\end{eqnarray}
The default values are $q_{_{\rm MIN}}=0.00$ and $\gamma=0.00$, i.e. a flat distribution with no minimum \citep[e.g.][]{Raghetal2010,Jansetal2012}.

Finite $q_{_{\rm MIN}}$ reflects the possibility that Sun-like primaries may eschew brown-dwarf or very low-mass secondaries (the so-called Brown Dwarf Desert). We therefore consider the range $0.00\leq q_{_{\rm MIN}}\leq 0.10$.

Regarding the range of $\gamma$, we adopt $-0.60<\gamma\leq 2.00$ \citep[see][and references therein]{DuchKrau2013}.

Negative $\gamma$ means a preference for low-mass companions, and $\gamma\!=\!-0.60$ means that (with $q_{_{\rm MIN}}\!=\!0$) $\sim\!50\%$ of systems have $q\!<\!0.2$. This might be appropriate for intermediate-mass and/or wide systems. However, the observational evidence for a predominance of such low $q$ values may actually refer to low-mass tertiary components orbiting unresolved binaries.

Positive $\gamma$ means a preference for a companion of comparable mass \citep[or ``twin", e.g.][]{Lucy2006}, and $\gamma\!=\!2.00$ means that (with $q_{_{\rm MIN}}\!=\!0$) $\sim\!50\%$ of systems have $q\!>\!0.8$. However, the observational evidence for a predominance of such high $q$ values may reflect selection effects. It is mainly seen in systems at the extremes of primary mass, where the statistical limitations of the observational data are most severe, and our concern here is not with such extreme masses. 

%%%%%%%%%%%%%%
\begin{table*}
\begin{center}
\begin{tabular}{|l|c|cc|cc|}\hline
{\sc Source} & $m_{_{\rm S}}\!\left({\rm M}_{_\odot}\right)$ & $\beta_{_0}$ & $\beta_{_1}$ & $b_{_\star}\!\left({\rm M}_{_\odot}\right)$ & $f_{_\star}\!\left({\rm M}_{_\odot}\right)$ \\\hline
\citet{Raghetal2010} & 0.44 & 0.50 & 0.46 & 0.535 & 0.461 \\
\citet{DuquMayo1991} & 0.58 & 0.64 & 0.52 & 0.667 & 0.450 \\\hline 
\end{tabular}
\caption{Column 1 gives the source of the estimate of the multiplicity frequency for systems with Sun-like primaries, $m_{_{\rm S}}\!\left({\rm M}_{_\odot}\right)$, and Column 2 gives the value of $m_{_{\rm S}}\!\left({\rm M}_{_\odot}\right)$. Columns 3 and 4 give the values of the parameters $\beta_{_0}$ and $\beta_{_1}$ in the expression for the probability that a system of given log-mass is a binary (Eqn. \ref{EQN:BINFRAC}). Columns 5 and 6 give the fraction of Sun-like stars that are in binaries, $b_{_\star}\!\left({\rm M}_{_\odot}\right)$, and the ratio of Sun-like secondaries to Sun-like primaries, $f_{_\star}\!\left({\rm M}_{_\odot}\right)$.}
\label{TAB:FITS}
\end{center}
\end{table*}
%%%%%%%%%%%%

%%%%%%%%%%%%%%%%%%%%%%%%%%%%%%%%%%%%%%%%%%%%%%%%%%%%%%%%%%%%%%%%%%%%%
\subsection{The distributions of singles, primaries and secondaries}%
%%%%%%%%%%%%%%%%%%%%%%%%%%%%%%%%%%%%%%%%%%%%%%%%%%%%%%%%%%%%%%%%%%%%%

With these definitions, the mass distribution of singles is
\begin{eqnarray}\label{EQN:SINGLES}
\frac{d{\cal N}}{d\mu_{_0}}&=&\left(\frac{d{\cal N}}{d\mu}\right)_{_{\mu\!=\!\mu_{_0}}}\,\left\{1-\beta(\mu_{_0})\right\}\,;
\end{eqnarray}
the mass distribution of primaries is
\begin{eqnarray}\nonumber
\frac{d{\cal N}}{d\mu_{_1}}&=&\int\limits_{\mu=\mu_{_1}}^{\mu=\mu_{_1}\!+\log_{_{10}}(2)}\,\frac{d{\cal N}}{d\mu}\,\beta(\mu)\\\nonumber
&&\hspace{2.0cm}\times\left(\frac{dP}{dq}\,\ell\,(q+1)\right)_{q=10^{(\mu-\mu_{_1})}-1}d\mu;\\\label{EQN:PRIMARIES}
\end{eqnarray}
and the mass distribution of secondaries is
\begin{eqnarray}\nonumber
\frac{d{\cal N}}{d\mu_{_2}}&=&\int\limits_{\mu=\mu_{_2}\!+\log_{_{10}}(2)}^{\mu=\infty}\,\frac{d{\cal N}}{d\mu}\,\beta(\mu)\\\nonumber
&&\hspace{1.25cm}\times\left(\frac{dP}{dq}\,\ell\;q(q+1)\right)_{q=\left(10^{(\mu-\mu_{_2})}-1\right)^{-1}}d\mu.\\\label{EQN:SECONDARIES}
\end{eqnarray}
The last terms in the integrands of Eqns. (\ref{EQN:PRIMARIES}) and (\ref{EQN:SECONDARIES}) represent, respectively, $|\partial P/\partial \mu|_{\mu_{_1}}$ and $|\partial P/\partial \mu|_{\mu_{_2}}$.

%%%%%%%%%%%%%%
\begin{figure}
\centering
\includegraphics[width=1.0\columnwidth]{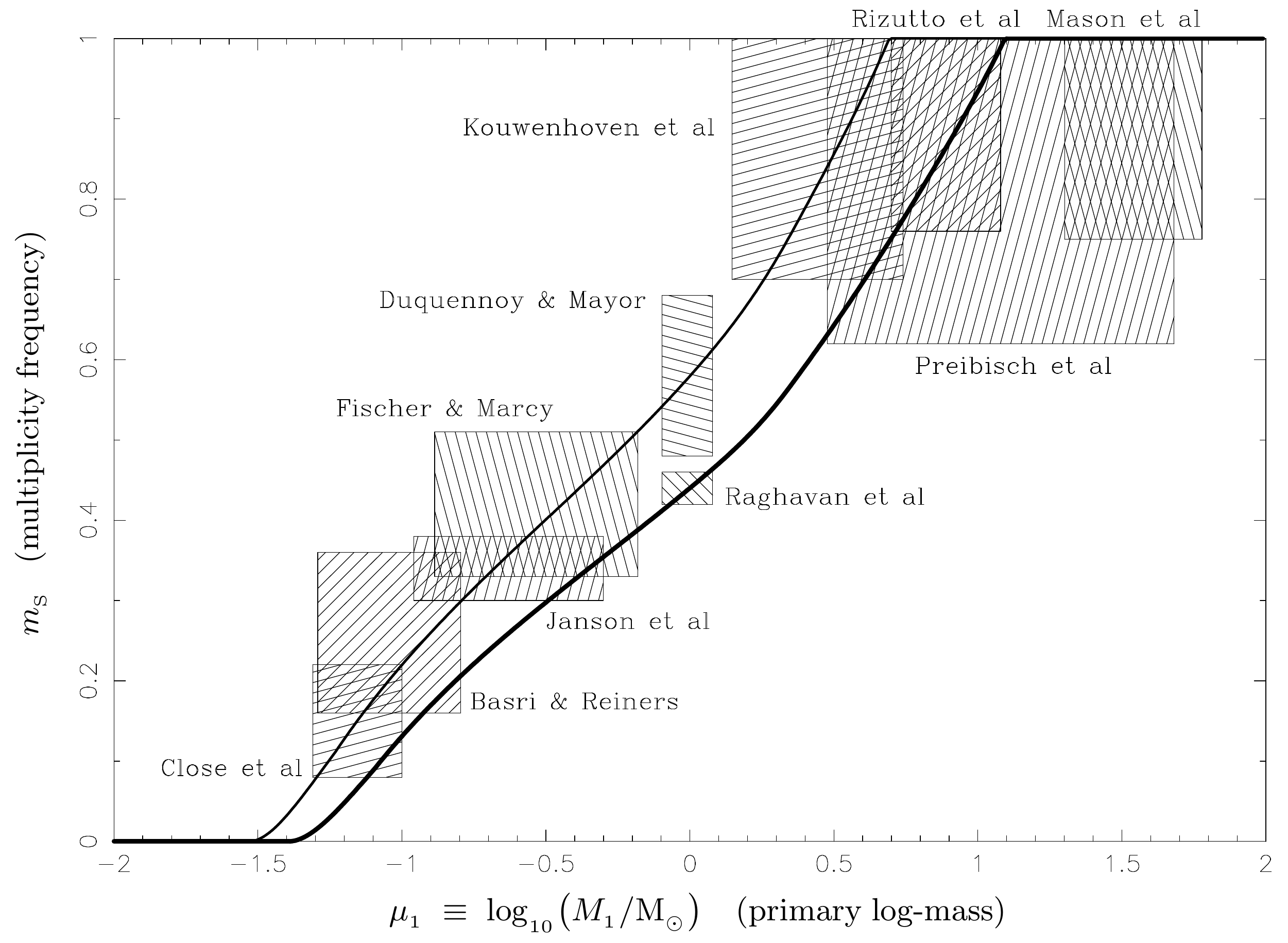}
\caption{The variation of multiplicity frequency, $m_{_{\rm S}}$, with primary log-mass, $\mu_{_1}\!\equiv\!\log_{_{10}}(M_{_1}/{\rm M}_{_\odot})$. The hatched acceptance boxes represent observational estimates from \citet{Closetal2003}, \citet{BasrRein2006}, \citet{FiscMarc1992}, \citet{Jansetal2012}, \citet{DuquMayo1991}, \citet{Kouwetal2007}, \citet{Rizzetal2013}, \citet{Preietal1999}, and \citet{Masoetal1998}; the width of each box represents the approximate range of masses considered, and the height of each box represents the range of $m_{_{\rm S}}$ deduced. Note that, for the higher mass ranges \citep[i.e.][]{Kouwetal2007, Rizzetal2013, Preietal1999, Masoetal1998}, there are only lower limits on $m_{_{\rm S}}$. The lower (bolder) line is the best fit when we require the model to go through the \citet{Raghetal2010} estimate of $m_{_{\rm S}}$ for systems with Sun-like primaries; this fit defines the default values of the parameters $\beta_{_0}$ and $\beta_{_1}$ (see Table \ref{TAB:PARAMS}). The upper (feinter) line is the best fit when we require the model to go through the \citet{DuquMayo1991} estimate of $m_{_{\rm S}}$ for systems with Sun-like primaries; evidently this line fits the other data more comfortably.}
\label{FIG:mfvsM1}
\end{figure}
%%%%%%%%%%%%

%%%%%%%%%%%%%%%%%%%%%%%%
\subsection{Statistics}%
%%%%%%%%%%%%%%%%%%%%%%%%

To obtain the binary statistics of stars having mass $M_\star$, we set $\mu_{_0}=\mu_{_1}=\mu_{_2}=\log_{_{10}}(M_\star/{\rm M}_{_\odot})$ in Eqns. (\ref{EQN:SINGLES}), (\ref{EQN:PRIMARIES}) and (\ref{EQN:SECONDARIES}), and evaluate the integrals numerically. The multiplicity fraction for systems having primaries of this mass is then
\begin{eqnarray}
m_{_{\rm S}}(M_\star)&=&\frac{d{\cal N}/d\mu_{_1}}{d{\cal N}/d\mu_{_0}+d{\cal N}/d\mu_{_1}}\,,
\end{eqnarray} 
but the fraction of such stars that are in binaries is
\begin{eqnarray}
b_\star(M_\star)&=&\frac{d{\cal N}/d\mu_{_1}+d{\cal N}/d\mu_{_2}}{d{\cal N}/d\mu_{_0}+d{\cal N}/d\mu_{_1}+d{\cal N}/d\mu_{_2}}\,.
\end{eqnarray}
We can also compute the ratio of stars of mass $M_\star$ that are secondaries, to stars of mass $M_\star$ that are primaries,
\begin{eqnarray}
f_\star(M_\star)&=&\frac{d{\cal N}/d\mu_{_2}}{d{\cal N}/d\mu_{_1}}\hspace{0.8cm}\left(=\;\frac{\left(b_\star-m_{_{\rm S}}\right)}{m_{_{\rm S}}\left(1-b_\star\right)}\right)\,;
\end{eqnarray}
this is the fractional increase in the number of stars of mass $M_\star$ in binaries that derives from taking account of secondaries. The subscripts $\star$ on $b$ and $f$ record that these are stellar properties.

%%%%%%%%%%%%%%%%%%%%%%%%%%%%%%%%%%%%%%%%%%%%%%%%%%%%%%%%%%%%%%%%%%%%%%%%
\subsection{Constraining $\beta_{_0}$ and $\beta_{_1}$}\label{SEC:FITb}%
%%%%%%%%%%%%%%%%%%%%%%%%%%%%%%%%%%%%%%%%%%%%%%%%%%%%%%%%%%%%%%%%%%%%%%%%

%%%%%%%%%%%%%%
\begin{table*}
\begin{center}
\begin{tabular}{|c|ccc|ccc|}\hline
$\stackrel{\mbox{\sc model}}{\mbox{\sc function}}$ & $\stackrel{\mbox{\sc model}}{\mbox{\sc parameter},\,X}$ & $\stackrel{\mbox{\sc default}}{\mbox{\sc value},\,X_{_{\rm O}}}$ & $\stackrel{\mbox{\sc range}}{\Delta X}$ & $\stackrel{dm_{_{\rm S}}}{\overline{dX}}$ & $\stackrel{db_\star}{\overline{dX}}$ & $\stackrel{df_\star}{\overline{dX}}$ \\\hline
{\sc distribution} & $\mu_{_{\rm S}}$ & $-0.60$ & $\pm 0.05$ & 0.13 & 0.18 & 0.27 \\
{\sc of system} & $\sigma_{_{\rm S}}$ & $+0.55$ & $\pm 0.05$ & 0.32 & 0.44 & 0.75 \\
{\sc log-masses} & $\alpha$ & $+2.35$ & $\pm 0.35$ & 0.00 & -0.02 & -0.14 \\\hline
{\sc binary} & $\beta_{_0}$ & $+0.50$ & $\pm 0.05$ & 0.90 & 0.90 & -0.20 \\
{\sc fraction} & $\beta_{_1}$ & $+0.46$ & $\pm 0.05$ & 0.07 & 0.09 & 0.15 \\\hline
{\sc distribution} & $q_{_{\rm MIN}}$ & $+0.00$ & $+0.10,\,-0.00$ & -0.07 & 0.03 & 0.59 \\
{\sc of mass ratios} & $\gamma$ & $+0.00$ & $+2.00,\,-0.60$ & -0.03 & 0.01 & 0.26 \\\hline
\end{tabular}
\caption{Column 1 gives the model parameters. Columns 2 through 4 give their symbols, their default values and their ranges. Columns 5 through 7 give the derivatives of, respectively, the multiplicity frequency of Sun-like stars, $m_{_{\rm S}}({\rm M}_{_\odot})$, the fraction of Sun-like stars that are in binaries, $b_\star({\rm M}_{_\odot})$, and the ratio of Sun-like secondaries to Sun-like primaries, $f_\star({\rm M}_{_\odot})$, with respect to each model parameter in turn; all the derivatives are evaluated for the default values of the model parameters.}
\label{TAB:PARAMS}
\end{center}
\end{table*}
%%%%%%%%%%%%

We identify the default values of $\beta_{_0}$ and $\beta_{_1}$ by requiring that the model (i) reproduce accurately the multiplicity frequency of Sun-like stars inferred from observation by \citet{Raghetal2010}, and (ii) reproduce as closely as possible the run of multiplicity frequency with primary mass derived from other observational studies by \citet{Closetal2003}, \citet{BasrRein2006}, \citet{FiscMarc1992}, \citet{Jansetal2012}, \citet{DuquMayo1991}, \citet{Kouwetal2007}, \citet{Rizzetal2013}, \citet{Preietal1999} and \citet{Masoetal1998}. The resulting default values are $\beta_{_0}\!=\!0.50$ and $\beta_{_1}\!=\!0.46$, so the probability that a system of mass $M$ is a binary becomes
\begin{eqnarray}
\beta(M)&=&\left\{\begin{array}{l}
0, \hspace{2.7cm} M\!<\!0.08\,{\rm M}_{_\odot};\\
0.50+0.46\log_{_{10}}\!\left(\!M/{\rm M}_{_\odot}\!\right)\!, \\
$\,$\hspace{1.5cm} 0.08\,{\rm M}_{_\odot}\!\leq\!M\!\leq\!12.5\,{\rm M}_{_\odot};\\
1, \hspace{2.7cm} M\!>\!12.5\,{\rm M}_{_\odot}.
\end{array}\right.
\end{eqnarray}
The corresponding fit to the observational estimates is illustrated by the lower (bolder) line on Fig. \ref{FIG:mfvsM1}. A brief discussion of this plot is appropriate.

The model equations and default parameters describing {\it both} the distribution of system masses (i.e. Chabrier log-normal at low and intermediate masses, with $\mu_{_{\rm S}}\!=\!-0.60$ and $\sigma_{_{\rm S}}\!=\!0.55$, plus Salpeter power law, with negative slope $\alpha\!=\!2.35$, at high masses), {\it and} the distribution of stellar mass ratios (flat, $\gamma\!=\!0.00$, with no minimum, $q_{_{\rm MIN}}\!=\!0.00$) appear to be the natural default choices; we will explore the consequences of varying the model parameters, both individually and collectively, in \S \ref{SEC:RES}. The justification for adopting Eqn. (\ref{EQN:BINFRAC}) for the fraction of systems that are binaries as a function of system mass, and hence the choice of the model parameters $\beta_{_0}$ and $\beta_{_1}$, is more {\it adhoc}.

From the observational data presented in Fig. \ref{FIG:mfvsM1}, it appears that, for $\,-1\la\,\mu_{_1}\la 1$, $\,m_{_{\rm S}}$ increases approximately linearly with $\mu_{_1}$, and hence a linear relation between $\beta$ and $\mu$ (i.e. Eqn. \ref{EQN:BINFRAC}) seems the simplest option to explore. However, it is also clear from Fig. \ref{FIG:mfvsM1} that Eqn. (\ref{EQN:BINFRAC}) gives a much better fit when the \citet{DuquMayo1991} acceptance box is invoked for systems with Sun-like primaries (upper, feinter line) than whan the \citet{Raghetal2010} acceptance box is invoked (lower, bolder line). Even if the \citet{FiscMarc1992} and \citet{DuquMayo1991} acceptance boxes are discounted, the \citet{Raghetal2010} acceptance box is hard to reconcile with the others without introducing a more complex function $m_{_{\rm S}}(\mu_{_1})$, for which the slope has a distinct minimum around $\mu_{_1}\!\sim\!1$. Since -- as far as we are aware -- no physical explanantion for such a minimum has been advanced, we avoid this complication. It would certainly be convenient if, when in future the multiplicity frequencies for systems with non--Sun-like primaries are evaluated more accurately, the values are reduced (say by $\sim 20\%$) to bring them into line with \citet{Raghetal2010}, {\it or} if the \citet{Raghetal2010} estimate is revised upwards.

In this context, it is important to note that Sun-like singles derive from systems with $\,\mu\!=\! 0$, Sun-like primaries from systems with $\,0\!<\!\mu\!\leq\log_{_{10}}(2)$, and Sun-like secondaries from systems with $\,\log_{_{10}}(2)\!<\!\mu\!<\!\infty$. In addition, with any sensible choice of the model parameters, the system mass function is falling quite rapdly with increasing $\mu$ for $\mu\!\geq\!0$. Thus the important range of applicability of Eqn. (\ref{EQN:BINFRAC}) is $\,0\!\leq\!\mu\!\la\!0.6$.

At low $\,\mu\;(\mu\leq\! -\beta_{_0}/\beta_{_1}$, $M<0.08\,{\rm M}_{_\odot}$), $\beta$ has to be set to zero to avoid non-physical predictions, and this has the consequence that the multiplicity frequency falls to zero for $\mu_{_1}\!\la\!-1.3$ (i.e. $M_{_1}\la 0.05\,{\rm M}_{_\odot}$). The observed multiplicity frequency appears to be very low for such low-mass primaries (only one of the systems reported by \citet{Closetal2003} has $M_{_1}\la 0.05\,{\rm M}_{_\odot}$), but it is probably not zero. We explore the effect of adopting model parameters that increase the multiplicity frequency of brown dwarfs and very low-mass stars in \S \ref{SEC:RES}. However, it is important to note that the value of $\beta$ at these low masses is irrelevant to the statistics of Sun-like stars, because a low-mass core, $M<{\rm M}_{_\odot}$, cannot spawn a Sun-like star.

At high $\,\mu\;(\mu\geq (1-\beta_{_0})/\beta_{_1}$, $M>12.5\,{\rm M}_{_\odot}$), $\beta$ has to be set to one to avoid non-physical consequences. In fact the acceptance boxes due to \citet{Kouwetal2007}, \citet{Rizzetal2013}, \citet{Preietal1999} and \citet{Masoetal1998} are all lower limits. At these high primary masses, the multiplicity frequency is an inadequate measure of multiplicity, because higher-order multiple systems become increasingly important at high primary mass, and the multiplicity frequency does not distinguish between higher-order multiples and binaries \citep[][]{HubbWhit2005}. The pairing factor (number of orbits per system) or companion frequency (mean number of companions) would be more appropriate measures. In addition, the notion of high-mass {\it field} stars is fraught, because the highest-mass stars do not live long enough for their birth clusters to disolve completely into the field, and therefore many high-mass stars in the field are runaways, which have been ejected in violent $N$-body interactions, and therefore seldom, if ever, have companions. The multiplicity statistics for high-mass stars, which point to very high pairing factors and companion frequencies, tend actually to pertain to stars that are still intimately involved with their birth clusters.

%%%%%%%%%%%%%%%%%%%%%%%%%%%%%%%%%
\section{Results}\label{SEC:RES}%
%%%%%%%%%%%%%%%%%%%%%%%%%%%%%%%%%

%%%%%%%%%%%%%%%%%%%%%%%%%%%%%%%%%%
\subsection{The default solution}%
%%%%%%%%%%%%%%%%%%%%%%%%%%%%%%%%%%

For Sun-like stars and the default parameter set, we obtain $m_{_{\rm S}}({\rm M}_{_\odot})\!=\!0.440$, $b_\star({\rm M}_{_\odot})\!=\!0.535$ and $f_\star({\rm M}_{_\odot})\!=\!0.42$. In other words, although for every $\sim\!56$ single Sun-like stars, there are only $\sim\!44$ Sun-like primaries in binary systems, there are also $\sim\!20$ Sun-like secondaries in binary systems, hence a total of $\sim\!66$ Sun-like stars in binaries.

%%%%%%%%%%%%%%%%%%%%%%%%%%%%%%%%%%%%%%%%%%%%%%%%%%%%%%%%
\subsection{The effect of varying the model parameters}%
%%%%%%%%%%%%%%%%%%%%%%%%%%%%%%%%%%%%%%%%%%%%%%%%%%%%%%%%

\begin{figure}
\centering
\includegraphics[width=1.0\columnwidth]{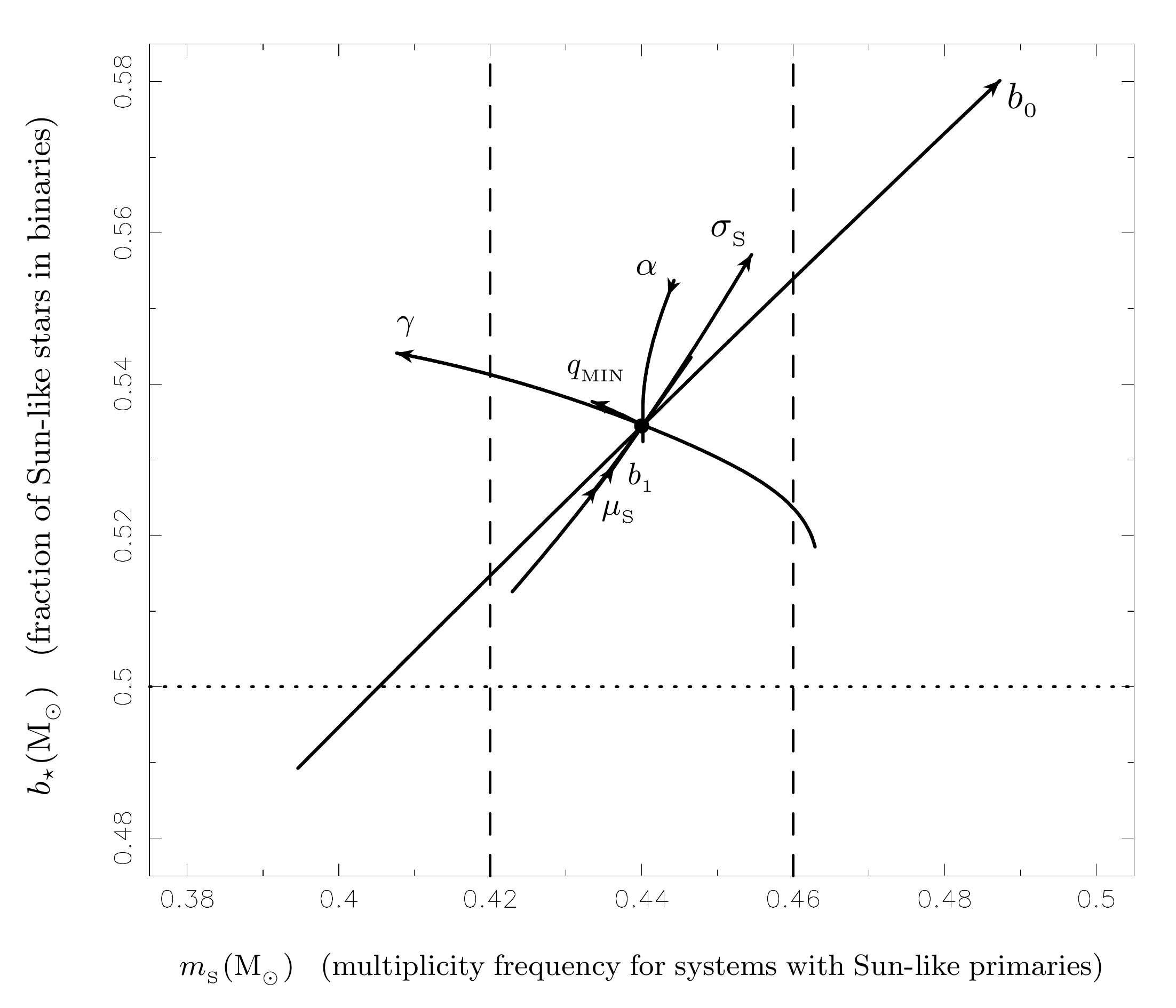}
\caption{The interdependence of the multiplicity frequency for systems having Sun-like primaries ($m_{_{\rm S}}({\rm M}_{_\odot})$; abscissa) and the fraction of Sun-like stars that are in binary systems ($b_\star({\rm M}_{_\odot})$; ordinate). The filled circle near the centre represents the combination obtained with the default parameters (see Table \ref{TAB:PARAMS}). The lines track the changes in $m_{_{\rm S}}({\rm M}_{_\odot})$ and $b_\star({\rm M}_{_\odot})$ that occur when the model parameters are changed individually between the limits in Table \ref{TAB:PARAMS} (for example, when $\mu_{_{\rm S}}$ varies from -0.65 to -0.55). In each case, an arrow at one end of the line indicates the direction in which the parameter increases, and the symbol for the parameter in question is always located beside this arrow. The lines for $\mu_{_{\rm S}}$, $\sigma_{_{\rm S}}$ and $\beta_{_1}$ are almost exactly on top of one another. Since each of these lines extends almost the same distance on either side of the default point, their lengths can be estimated from the locations of the corresponding arrows; the longest line is for $\sigma_{_{\rm S}}$ and the shortest for $\beta_{_1}$. The two vertical dashed lines mark the range of $m_{_{\rm S}}({\rm M}_{_\odot})$ deduced by \citet{Raghetal2010}. The horizontal dotted line is at $b_\star({\rm M}_{_\odot})\!=\!0.500$, so all solutions abve this line represent situations in which the majority of Sun-like stars are in multiple systems. We see that no individual parameter can reduce $b_\star({\rm M}_{_\odot})$ below this line without also taking $m_{_{\rm S}}({\rm M}_{_\odot})$ outside the range derived by \citet{Raghetal2010}.}
\label{FIG:Variance}
\end{figure}

\begin{figure}
\centering
\includegraphics[width=1.0\columnwidth]{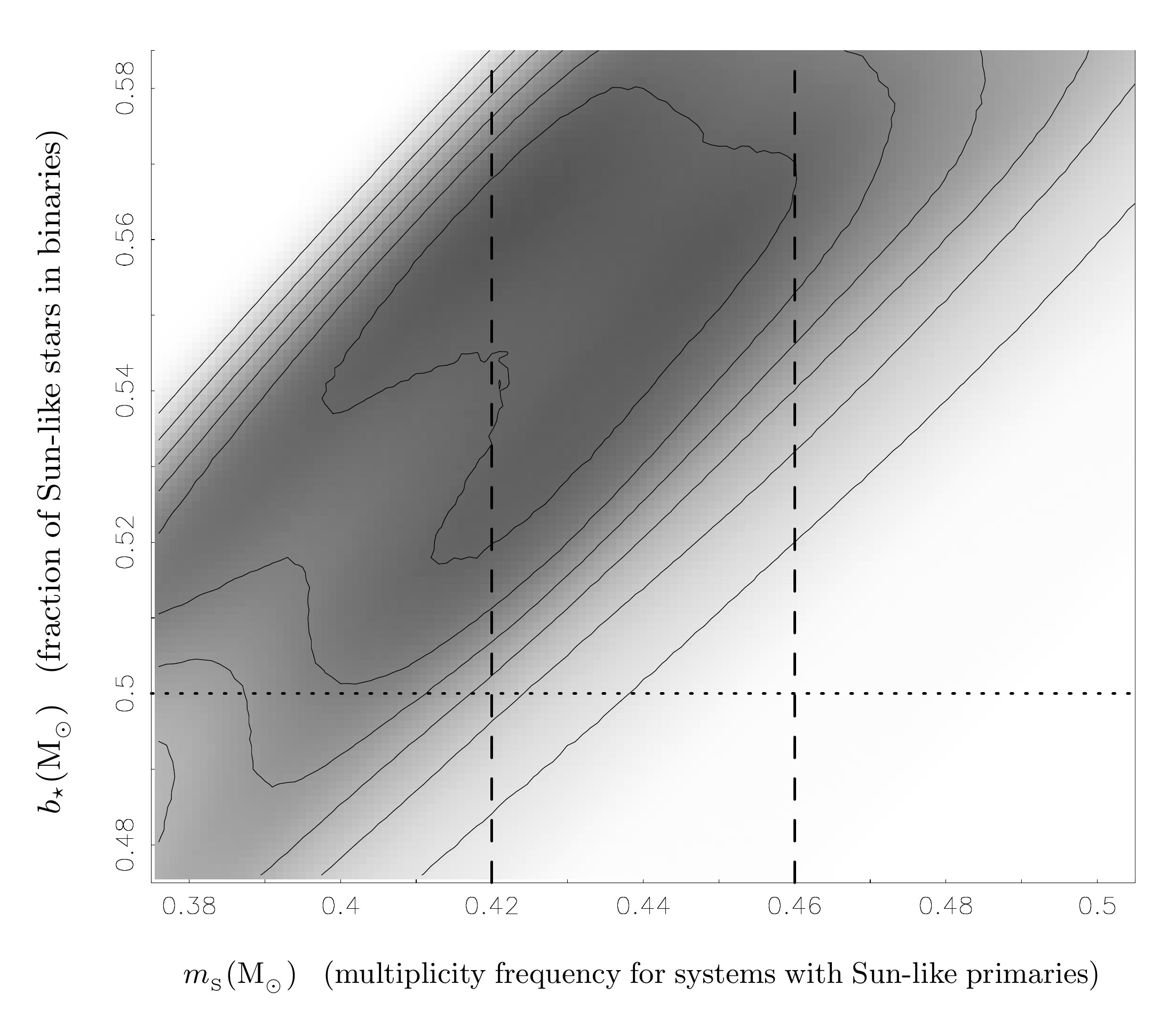}
\caption{The likelihood of different combinations of $m_{_{\rm S}}({\rm M}_{_\odot})$ (the multiplicity frequency for systems having Sun-like primaries; abscissa) and $b_\star({\rm M}_{_\odot})$ (the fraction of Sun-like stars that are in binary systems; ordinate). Different random sets of model parameters have been generated by picking a value for each parameter from a Gaussian distribution having mean equal to the default value and standard deviation equal to the specified range; thus, for any parameter there is a $\sim 32\%$ chance that it falls outside the range $X_{_{\rm O}}\pm\Delta X$. The grey-scale and contours delineate the relative probability of different combinations of $m_{_{\rm S}}({\rm M}_{_\odot})$ and $b_\star({\rm M}_{_\odot})$ that fall within the bounds of the figure. The contours are at $15\%$, $30\%$, ... $90\%$ of the peak probability. If we consider all the combinations generated (including those that fall outside the bounds of the figure), there is a $\sim\!21\%$ chance that the majority of Sun-like stars are single. However, this result is dominated by parameter sets with very low $\beta_{_1}$ which also deliver $m_{_{\rm S}}({\rm M}_{_\odot})$ well below the minimum estimated by \citet{Raghetal2010}, i.e. the extension into the bottom lefthand corner of the figure. If we limit our consideration to the sets that deliver $m_{_{\rm S}}({\rm M}_{_\odot})$ in the range deduced by \citet{Raghetal2010}, the chance that the majority of Sun-like stars are single falls to $\sim\!3\%$.}
\label{FIG:MonteCarlo}
\end{figure}

Figs. \ref{FIG:Variance} and \ref{FIG:MonteCarlo} represent the $(m_{_{\rm S}}({\rm M}_{_\odot}),b_\star({\rm M}_{_\odot}))$-plane in the vicinity of the solution obtained with the default parameters (hereafter the default solution). The black dot at $m_{_{\rm S}}({\rm M}_{_\odot})\!=\!0.440$ and $b_\star({\rm M}_{_\odot})\!=\!0.535$ on Fig. \ref{FIG:Variance} represents the default solution. The vertical dashed lines demark the limits on $m_{_{\rm S}}({\rm M}_{_\odot})$ obtained by \citet{Raghetal2010}, viz. $0.42\!<\!m_{_{\rm S}}({\rm M}_{_\odot})\!<\!0.46$. The horizontal dotted line is at $b_\star({\rm M}_{_\odot})\!=\!0.50$; for solutions below (above) this line more (less) than $50\%$ of Sun-like stars are single.

On Fig. \ref{FIG:Variance}, the lines passing through the default solution show how the solution changes if one of the model parameters is varied. The arrow at one end of a line indicates the direction in which the parameter increases, and the symbol for the parameter in question is given beside this arrow. The lines for $\mu_{_{\rm S}}$, $\sigma_{_{\rm S}}$ and $\beta_{_1}$ lie almost on top of one another; each of these lines extends approximately the same distance on either side of the default solution, and therefore the length can be inferred from the position of the corresponding arrow (the line for $\sigma_{_{\rm S}}$ is the longest, and that for $\beta_{_1}$ the shortest).

Over the range of system log-masses that contributes Sun-like stars, $0\!\leq\!\mu\!<\!\infty$, the mass function is decreasing with increasing $\mu$, i.e. $d{\cal N}/d\mu\!<\!0$. Increasing $\mu_{_{\rm S}}$ and/or increasing $\sigma_{_{\rm S}}$ and/or decreasing $\alpha$ increases $d{\cal N}/d\mu\!<\!0$, i.e. {\it reduces} the downward slope of the system mass function, so there are then more Sun-like primaries relative to singles (increased $m_{_{\rm S}}({\rm M}_{_\odot})$) and more Sun-like secondaries relative to primaries (increased $b_\star({\rm M}_{_\odot}))$. Conversely, decreasing $\mu_{_{\rm S}}$ and/or decreasing $\sigma_{_{\rm S}}$ and/or increasing $\alpha$ decreases $m_{_{\rm S}}({\rm M}_{_\odot})$ and $b_\star({\rm M}_{_\odot})$. Increasing $\alpha$ from its minimum value ($\alpha\!=\!2.00$) has an ever diminishing effect, because the switch from the log-normal system mass distribution to the power-law distribution occurs at ever increasing $\mu_{_{\rm C}}$ (see Eqn. \ref{EQN:muC}); consequently the line for increasing $\alpha$ does not extend far beyond the default solution on Fig. \ref{FIG:Variance}, once $\alpha\!>\!2.35$.

Increasing $\beta_{_0}$ and/or $\beta_{_1}$ above the default values increases the number of primaries and secondaries, relative to singles, and hence increases both $m_{_{\rm S}}({\rm M}_{_\odot})$ and $b_\star({\rm M}_{_\odot})$; increasing $\beta_{_0}$ has a larger effect than increasing $\beta_{_1}$ because the binary statistics of Sun-like stars are dominated by systems with $0\!\leq\!\mu\!\la\!0.6$, i.e. relatively small $\mu$. Conversely, decreasing $\beta_{_0}$ and/or $\beta_{_1}$ below their default values decreases both $m_{_{\rm S}}({\rm M}_{_\odot})$ and $b_\star({\rm M}_{_\odot})$.

Increasing $q_{_{\rm MIN}}$ and/or $\gamma$ above the default values squeezes the distribution of systems involving Sun-like primaries towards the upper end of the available range ($0\!<\!\mu\!\leq\!\log_{_{10}}(2)$), i.e. {\it down} the falling system log-mass distribution, thereby reducing the number of Sun-like primaries. At the same time, it squeezes  the distribution of systems involving Sun-like secondaries towards the lower end of the available range ($\log_{_{10}}(2)\!<\!\mu\!\leq\!\infty$), i.e. {\it up} the falling system log-mass distribution, thereby augmenting the number of Sun-like secondaries. Consequently $m_{_{\rm S}}({\rm M}_{_\odot})$ is reduced, but $b_\star({\rm M}_{_\odot})$ is augmented. $q_{_{\rm MIN}}$ cannot be reduced below the default value, but decreasing $\gamma$ below the default value augments $m_{_{\rm S}}({\rm M}_{_\odot})$ and reduces $b_\star({\rm M}_{_\odot})$.

Table \ref{TAB:PARAMS} gives the parameter symbols; their default values and prescribed ranges; and the partial derivatives, $\partial m_{_{\rm S}}({\rm M}_{_\odot})/\partial X$, $\partial b_\star({\rm M}_{_\odot})/\partial X$ and $\partial f_\star({\rm M}_{_\odot}) /\partial X$, where $X$ is one of the model parameters (i.e. $X\equiv\mu_{_{\rm S}},\;\sigma_{_{\rm S}},\;\beta_{_0},\;\beta_{_1},\;q_{_{\rm MIN}},\;\gamma$), and all the derivatives are evaluated for the default parameter set.

Fig. \ref{FIG:MonteCarlo} displays the distribution of solutions when the model parameters are varied simultaneously and randomly. This plot is produced by generating $\sim\!1.8\times 10^9$ different solutions ($\sim\!10^9$ of which fall within Fig. \ref{FIG:MonteCarlo}). For each solution, the value of parameter $X$ is chosen by generating a random Gaussian deviate, ${\cal G}$, from a distribution with mean $0$ and standard deviation 1, and then putting $X\!=\!X_{_{\rm O}}\!+\!{\cal G}\,\Delta\! X$, where the values of $X_{_{\rm O}}$ and $\Delta X$ for each model parameter are given in Table \ref{TAB:PARAMS} (Columns 3 and 4); with this procedure, $\sim 32\%$ of parameter values fall outside the range $X_{_{\rm O}}\pm\Delta\!X$.

If we consider all the solutions generated in this way (including those that fall outside Fig. \ref{FIG:MonteCarlo}), $\sim 21\%$ of them give $f_\star\!<\!0.50$. However, the majority of these solutions involve very low values of $\beta_{_0}$, and therefore they also deliver $m_{_{\rm S}}({\rm M}_{_\odot})\!<\!0.42$ \citep[i.e. below the range estimated by][]{Raghetal2010}.

If we limit consideration to solutions that satisfy the constraints calculated by \citet{Raghetal2010}, i.e. $0.42\!\leq\!m_{_{\rm S}}({\rm M}_{_\odot})\!\leq\!0.46$ (including those that fall outside Fig. \ref{FIG:MonteCarlo}), then $\sim 3\%$ of the allowed solutions give $f_\star\!<\!0.50$, because these solutions require both very low $\beta_{_0}$ and very low $\gamma$. We conclude that it is rather unlikely that the majority of Sun-like stars are single.

%%%%%%%%%%%%%%
%%%%%%%%%%%%

%%%%%%%%%%%%%%
%%%%%%%%%%%%

%%%%%%%%%%%%%%%%%%%%%%%%%%
\subsection{Other masses}%
%%%%%%%%%%%%%%%%%%%%%%%%%%

If we accept the default parameters as a reasonable representation of the binary statistics of low- and intermediate-mass stars, we can estimate the fraction of such stars that are single. For example, for $M_\star\!\leq\!0.70\,{\rm M}_{_\odot}$, $\geq\!50\%$ of stars are single. Thus most M Dwarfs are single.

%%%%%%%%%%%%%%%%%%%%%%%%%%%%%%%%%%%%%%
\section{Conclusions}\label{SEC:CONC}%
%%%%%%%%%%%%%%%%%%%%%%%%%%%%%%%%%%%%%%

We have developed a model to estimate the fraction of Sun-like stars that are secondaries in binary systems. Our model (which only considers singles and binary systems, and therefore gives a lower limit to the fraction of Sun-like stars that are in multiples) invokes a log-normal distribution of system masses \citep[][]{Chabrier2005} with a power-law tail at high masses \citep[][]{Salpeter1955}; a flat distribution of mass ratios \citep[e.g.][]{Raghetal2010, Jansetal2012, ReggMeye2013}; and the probability that a system with mass $M$ is a binary,
\[\beta\!=\!0.50+0.47\log_{_{10}}(M/{\rm M}_{_\odot})\,.\]
We find that, even if the multiplicity frequency is as low as the recent estimate of \citet{Raghetal2010}, $0.42\!\leq\!m_{_{\rm S}}({\rm M}_{_\odot})\!\leq\!0.46$, and even if we vary the model parameters significantly from their default values, when account is taken of secondaries (and higher-order components in mutiple systems), the majority of Sun-like stars are probably in multiple systems. Our model predicts that the fraction of stars that are in multiple systems is a monotonically increasing function of stellar mass, and that the majority of stars with mass $M_\star\!>\!0.7\,{\rm M}_{_\odot}$ are in multiple systems. Conversely, most M Dwarfs, and hence most stars overall, are single.

\section*{Acknowledgements}

APW and OL gratefully acknowledge the support of a consolidated grant (ST/K00926/1) from the UK STFC. We thank the referee, Mike Simon, for his useful comments.

%%%%%%%%%%%%%%%%%%%%%%%%
\bibliographystyle{mn2e}
\bibliography{antsrefs}
%%%%%%%%%%%%%%%%%%%%%%%

\label{lastpage}
\end{document}